\documentclass[aps, showpacs, showkeys,nofootinbib,floatfix]{revtex4}

\usepackage{amssymb}
\usepackage{amsmath}
\usepackage{graphicx}

\begin{document}

\title{Holographic Dark Energy in Brans-Dicke Theory}

\author{Lixin Xu\footnote{Corresponding author}}
\email{lxxu@dlut.edu.cn}
\author{Wenbo Li}
\author{Jianbo Lu}

\affiliation{Institute of Theoretical Physics, School of Physics \&
Optoelectronic Technology, Dalian University of Technology, Dalian,
116024, P. R. China}

\begin{abstract}
In this paper, the holographic dark energy model is considered in
Brans-Dicke theory where the holographic dark energy density
$\rho_{\Lambda} =3c^2 M^{2}_{pl} L^{-2}$ is replaced with
$\rho_{h}=3c^2 \Phi(t)L^{-2}$. Here $\Phi(t)=\frac{1}{8\pi G}$ is a
time variable Newton constant. With this replacement, it is found
that no accelerated expansion universe will be achieved when the
Hubble horizon is taken as the role of IR cut-off. When the event
horizon is adopted as the IR cut-off, an accelerated expansion
universe is obtained. In this case, the equation of state of
holographic dark energy $w_h$ takes a modified form
$w_h=-\frac{1}{3}\left(1+\alpha+\frac{2}{c}\sqrt{\Omega_{h}}\right)$.
In the limit $\alpha\rightarrow 0$, the 'standard' holographic dark
energy is recovered. In the holographic dark energy dominated epoch,
power-law and de Sitter time-space solutions are obtained.
\end{abstract}


\hfill ITP-DUT/2008-08

\maketitle

\section{Introduction}
The observation of the Supernovae of type Ia
\cite{ref:Riess98,ref:Perlmuter99} provides the evidence that the
universe is undergoing accelerated expansion. Jointing the
observations from Cosmic Background Radiation
\cite{ref:Spergel03,ref:Spergel06} and SDSS
\cite{ref:Tegmark1,ref:Tegmark2}, one concludes that the universe at
present is dominated by $70\%$ exotic component, dubbed dark energy,
which has negative pressure and push the universe to accelerated
expansion. Of course, the accelerated expansion can attribute to the
cosmological constant naturally. However, it suffers the so-called
{\it fine tuning} and {\it cosmic coincidence} problem. To avoid
these problem, dynamic dark energy models are considered, such as
quintessence
\cite{ref:quintessence1,ref:quintessence2,ref:quintessence3,ref:quintessence4},
phtantom \cite{ref:phantom}, quintom \cite{ref:quintom} and
holographic dark energy \cite{ref:holo1,ref:holo2} etc. To explain
the accelerated expansion, modified gravity theories are explored
too. For recent reviews, please see
\cite{ref:DEReview1,ref:DEReview2,ref:DEReview3,ref:DEReview4,ref:DEReview5,ref:DEReview6}.
Brans-Dicke theory \cite{ref:BransDicke} as a natural extension of
Einstein's general theory of relativity can pass the experimental
tests from the solar system \cite{ref:solar} and provide explanation
to the accelerated expansion of the universe
\cite{ref:BDEp1,ref:BDEp2,ref:BDEp3}. In Brans-Dicke theory, the
gravitational constant is replaced with a inverse of time dependent
scalar field, i.e. $8\pi G=\frac{1}{\Phi(t)}$, which couples to
gravity with a coupling parameter $\omega$.

Recently, a model named holographic dark energy has been discussed
extensively. The model is constructed by considering the holographic
principle and some features of quantum gravity theory. According to
the holographic principle, the number of degrees of freedom in a
bounded system should be finite and has relations with the area of
its boundary. By applying the principle to cosmology, one can obtain
the upper bound of the entropy contained in the universe. For a
system with size $L$ and UV cut-off $\Lambda$ without decaying into
a black hole, it is required that the total energy in a region of
size $L$ should not exceed the mass of a black hole of the same
size, thus $L^3\rho_{\Lambda} \le L M^2_{pl}$. The largest $L$
allowed is the one saturating this inequality, thus $\rho_{\Lambda}
=3c^2 M^{2}_{pl} L^{-2}$, where $c$ is a numerical constant and
$M_{pl}$ is the reduced Planck Mass $M^{-2}_{pl}=8 \pi G$.
It just means a {\it duality} between UV cut-off and IR cut-off. The
UV cut-off is related to the vacuum energy, and IR cut-off is
related to the large scale of the universe, for example Hubble
horizon, event horizon or particle horizon as discussed by
\cite{ref:holo1,ref:holo2}. In the paper \cite{ref:holo2}, the
author takes the future event horizon
\begin{equation}
R_{eh}(a)=a\int^{\infty}_{t}\frac{dt^{'}}{a(t^{'})}=a\int^{\infty}_{a}\frac{da^{'}}{Ha^{'2}}\label{eq:EH}
\end{equation}
as the IR cut-off $L$. This horizon is the boundary of the volume a
fixed observer may eventually observe. One is to formulate a theory
regarding a fixed observer within this horizon. As pointed out in
\cite{ref:holo2}, it can reveal the dynamic nature of the vacuum
energy and provide a solution to the {\it fine tuning} and {\it
cosmic coincidence} problem. In this model, the value of parameter
$c$ determines the property of holographic dark energy. When $c\ge
1$, $c=1$ and $c\le 1$, the holographic dark energy behaviors like
quintessence, cosmological constant and phantom respectively.

\section{Friedmann Equation in Brans-Dicke Theory}
In this paper, we generalized the holographic dark energy model to
that in the framework of Brans-Dicke theory which has already been
considered by many authors
\cite{ref:BDH1,ref:BDH2,ref:BDH3,ref:BDH4,ref:BDH5}. Then, it takes
the general form
\begin{equation}
\rho_{h}=3c^2 \Phi(t)L^{-2},
\end{equation}
where $\Phi(t)=\frac{1}{8 \pi G}$ is a reverse of time variable
Newton constant. In a spatially flat FRW cosmology filled dark
matter and holographic dark energy, the gravitational equations can
be written as
\begin{eqnarray}
3\Phi\left[H^2+H\frac{\dot{\Phi}}{\Phi}-\frac{\omega}{6}\frac{\dot{\Phi}^2}{\Phi^2}\right]=\rho_{m}+\rho_{h},\label{eq:FE}\\
2\frac{\ddot{a}}{a}+H^2+\frac{\omega}{2}\frac{\dot{\Phi}^2}{\Phi^2}+2H\frac{\dot{\Phi}}{\Phi}+\frac{\ddot{\Phi}}{\Phi}=-\frac{p_{h}}{\Phi},\label{eq:RE}
\end{eqnarray}
where $H=\frac{\dot{a}}{a}$ is the Hubble parameter, $\rho_{m}$ is
dark matter energy density, $\rho_{h}$ is the holographic dark
energy density and $p_{h}$ is the pressure of holographic dark
energy. The scalar field evolution equation is
\begin{equation}
\ddot{\Phi}+3H\dot{\Phi}=\frac{\rho_{m}+\rho_{h}-3p_{h}}{2\omega+3}.\label{eq:phiE}
\end{equation}
Considering the dark matter energy conservation equation
\begin{equation}
\dot{\rho_{m}}+3H\rho_{m}=0,
\end{equation}
and jointing it with Eq. (\ref{eq:FE}), Eq. (\ref{eq:RE}) and Eq.
(\ref{eq:phiE}), one obtains the holographic dark energy
conservation equation
\begin{equation}
\dot{\rho_{h}}+3H(\rho_{h}+p_{h})=0.\label{eq:DECE}
\end{equation}
Here, we have considered non-interacting cases. The Friedmann
equation (\ref{eq:FE}) is
\begin{equation}
H^2=\frac{\rho_{m}+\rho_{h}}{3\Phi}-H\frac{\dot{\Phi}}{\Phi}+\frac{\omega}{6}\frac{\dot{\Phi}^2}{\Phi^2}.
\label{eq:SFE}
\end{equation}
With the assumption $\Phi/\Phi_0=(a/a_0)^{\alpha}$, the Eq.
(\ref{eq:SFE}) is rewritten as
\begin{equation}
H^2=\frac{2}{(6+6\alpha-\omega\alpha^2)\Phi}(\rho_{m}+\rho_{h}).\label{eq:RFE}
\end{equation}
It is easy to find out that, in the limit case $\alpha \rightarrow
0$, the standard cosmology is recovered. To make the Friedmann
equation (\ref{eq:RFE}) to have physical meanings, i.e. to make
$(6+6\alpha-\omega\alpha^2)>0$, one has the following constraints on
the values of $\alpha$
\begin{equation}
\begin{array}{ccc}
\frac{3-\sqrt{9+6\omega}}{\omega}<\alpha<\frac{3+\sqrt{9+6\omega}}{\omega},
& \text{for} & \omega>0,\\
\alpha<\frac{3-\sqrt{9+6\omega}}{\omega} \quad \text{or} \quad
\alpha>\frac{3+\sqrt{9+6\omega}}{\omega},& \text{for} & -3/2
\leq \omega<0,\\
\Re, & \text{for} & \omega < -3/2.
\end{array}\label{eq:condition}
\end{equation}
However, the solar system experiments predict the value of $\omega$
is $\left|\omega\right| > 40000$ \cite{ref:solar}. However, the
value of parameter $\omega =-3/2$ is a boundary of ghost
\cite{ref:BDghost}. So, in this paper, when considering these
constraints, the second line of Eq. (\ref{eq:condition}) will be
omitted and $\omega> 40000$ will be consider in this paper. In fact,
authors \cite{ref:BDCos} have used the cosmic observations to
constrain the parameter $\omega$. In \cite{ref:BDCos}, the authors
found that $\omega$ can be smaller than $40000$ in cosmological
scale.

\section{Hubble Horizon as IR Cut-off}
At first, we consider the Hubble horizon as the IR cut-off, i.e.
$L=H^{-1}$. Then, the holographic dark energy is rewritten as
\begin{eqnarray}
\rho_{h}&=&3c^2\Phi H^2. \label{eq:Hde}
\end{eqnarray}
Inserting Eq.(\ref{eq:Hde}) into Eq. (\ref{eq:RFE}), one has
\begin{equation}
H^2=H^2_0
\frac{\Omega_{m0}}{1-\tilde{\Omega}_{h0}}(\frac{a_0}{a})^{3+\alpha},\label{eq:HFE}
\end{equation}
where $\Omega_{m0}=\frac{2}{(6+6\alpha-\omega
\alpha^2)}\frac{\rho_{m0}}{\Phi_0 H^2_0}$ and
$\tilde{\Omega}_{h0}=\frac{6c^2}{6+6\alpha-\omega \alpha^2}$. Then,
the equation (\ref{eq:HFE}) has the solution
\begin{equation}
a(t)=\left[\frac{3+\alpha}{2}\frac{\Omega_{m0}H^2_0}{1-\tilde{\Omega}_{h0}}\right]^{\frac{2}{3+\alpha}}a_0t^{\frac{2}{3+\alpha}}.
\end{equation}
To have an accelerated expansion, the condition
$\frac{2}{3+\alpha}>1$ is required, i.e. $\alpha<-1$. By combining
Eq. (\ref{eq:condition}) and taking the solar system constraint into
account, one can only take the value of $\omega <-40000$ to make an
accelerated expansion of the universe in Brans-Dicke theory when the
Hubble horizon is taken as the IR cut-off. However, if one considers
the current value of $\Phi_0=1/8\pi G$ and lets
$\Omega_{m0}=\frac{2}{(6+6\alpha-\omega
\alpha^2)}\frac{\rho_{m0}}{\Phi_0 H^2_0}\equiv\frac{8\pi
G\rho_{m0}}{3H^2_0}$, the relations $\omega\alpha=6$ and
$\tilde{\Omega}_{h0}=c^2$ will be derived. Then, combining with
solar system constraint $\left|\omega\right|
>40000$ and avoidance of ghost, one obtains
\begin{equation}
0<\alpha<\frac{3}{20000}.
\end{equation}
So, under this stronger condition, no accelerated expansion universe
will be achieved in Brans-Dicke theory without interactions, when
the Hubble horizon is taken as the IR cut-off.

\section{Event Horizon as IR Cut-off}
\begin{figure}[tbh]
\centering
\includegraphics[width=5.0in]{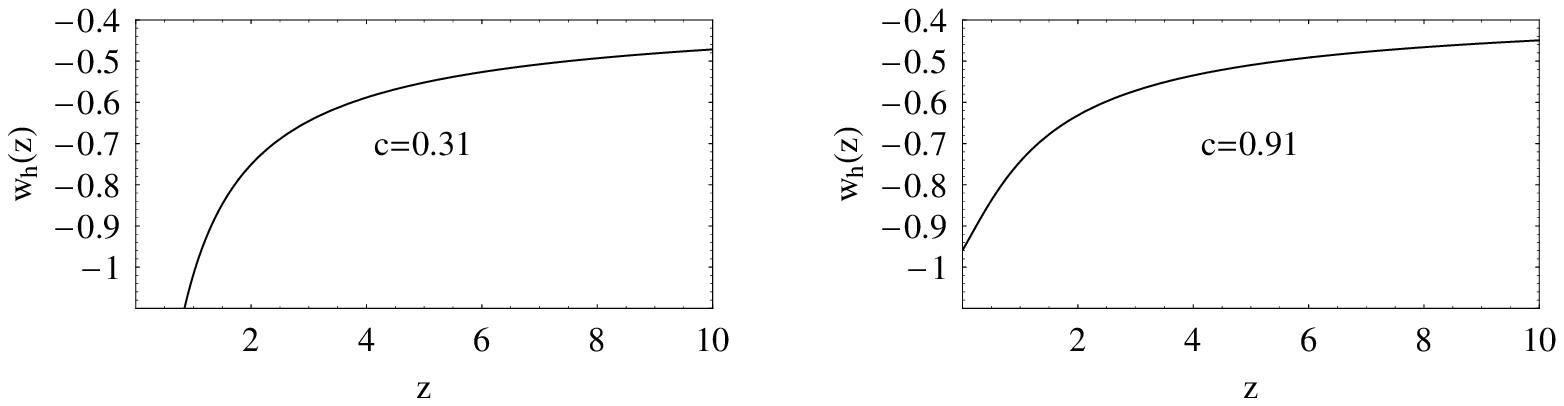}
\includegraphics[width=5.0in]{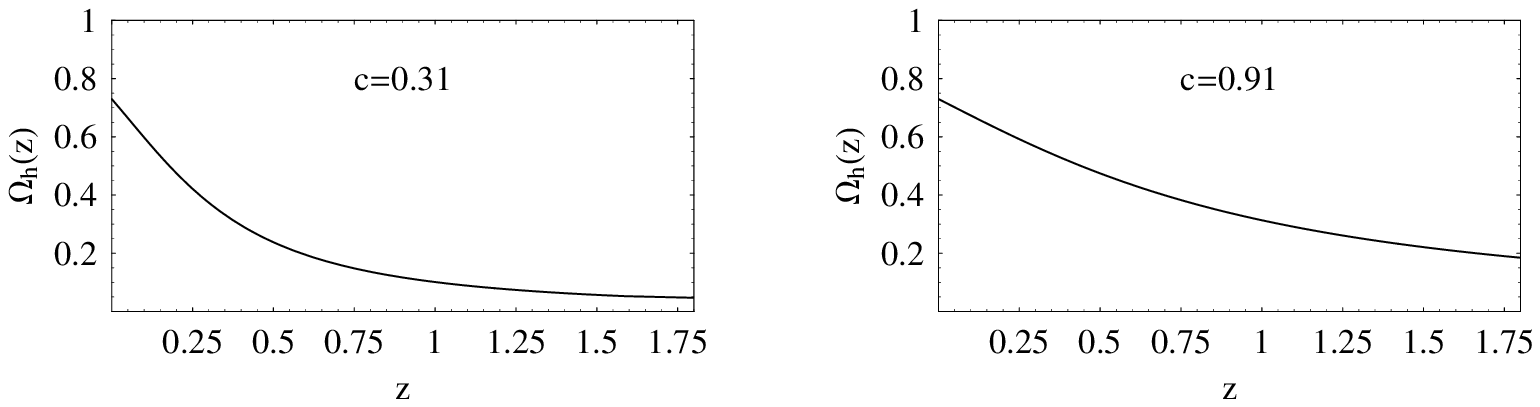}
\includegraphics[width=5.0in]{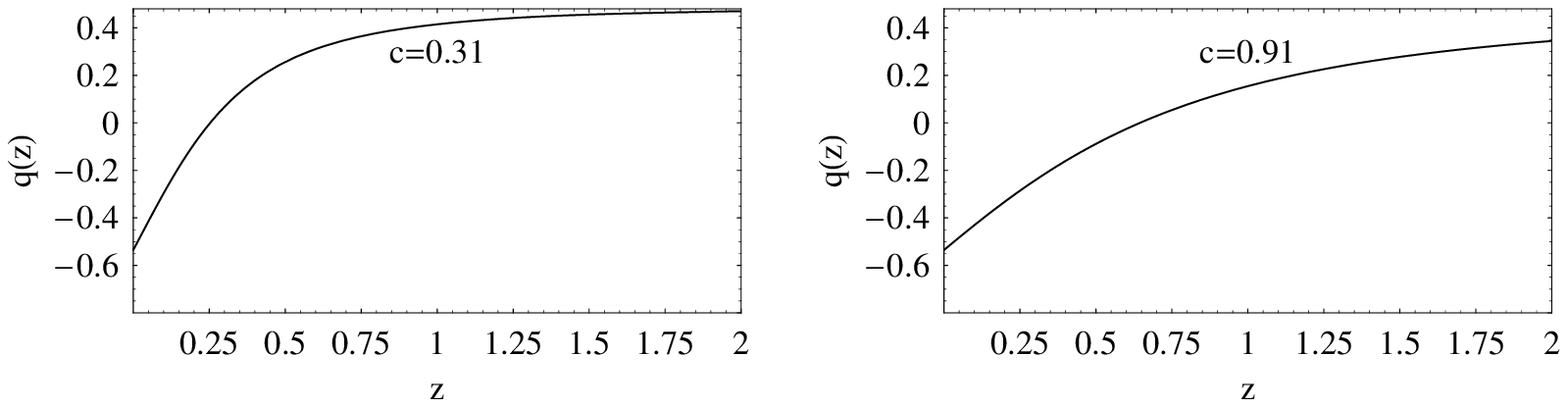}
\caption{The evolutions of EoS $w_h(z)$, dimensionless density
parameter $\Omega_{h}(z)$ of the holographic dark energy and the
deceleration parameter $q(z)$ with respect to the redshift $z$ in
Brans-Dicke theory, where the values $\Omega_{h0}=0.73$,
$c=0.31(c=0.91)$ and $\alpha=0.00005$ are adopted.}\label{fig:EOS}
\end{figure}

Now, as done in \cite{ref:holo2}, the event horizon $R_{eh}$ is
taken as the IR cut-off. Then, the holographic dark energy is
\begin{equation}
\rho_{h}=\frac{3c^2\Phi}{R^2_{eh}}.\label{eq:EHHDE}
\end{equation}
And, the Friedmann Eq. (\ref{eq:RFE}) is rewritten as
\begin{eqnarray}
H^2&=&H^2_0\Omega_{m0}\left(\frac{a_0}{a}\right)^{(3+\alpha)}+\Omega_{h}H^2
\nonumber \\
&=&H^2_0\Omega_{m0}a^{-(3+\alpha)}+\Omega_{h}H^2,\label{eq:EHFE}
\end{eqnarray}
where $\Omega_{h}=\frac{2}{6+6\alpha-\omega
\alpha^2}\frac{1}{\Phi}\frac{\rho_{h}}{H^2}=\tilde{\Omega}_{h0}\frac{1}{H^2
R^2_{eh}}$. For convenience, the scale factor $a$ has been
normalized to $a_0=1$. Jointing Eq. (\ref{eq:EHHDE}) and Eq.
(\ref{eq:EH}), one has
\begin{equation}
\int^{\infty}_{a}\frac{d\ln
a'}{Ha'}=\frac{1}{aH}\sqrt{\frac{\tilde{\Omega}_{h0}}{\Omega_{h}}}.\label{eq:re}
\end{equation}
From Eq.(\ref{eq:EHFE}), one obtains
\begin{equation}
\frac{1}{Ha}=\sqrt{a^{(1+\alpha)}(1-\Omega_{h})}\frac{1}{H_0
\sqrt{\Omega_{m0}}}.
\end{equation}
Inserting the above equation into Eq. (\ref{eq:re}), one has
\begin{equation}
\int^{\infty}_{x}e^{\frac{(1+\alpha)}{2}x'}\sqrt{1-\Omega_{h}}dx'=e^{\frac{(1+\alpha)}{2}x}\sqrt{\tilde{\Omega}_{h0}}\sqrt{\frac{1}{\Omega_{h}}-1},
\end{equation}
where $x=\ln a$. Taking derivative with respect to $x =\ln a$ from
both sides of the above equation, one has the differential equation
of $\Omega_{h}$
\begin{equation}
\Omega_{h}'=\Omega_{h}\left(1-\Omega_{h}\right)\left(1+\alpha+\frac{2}{\sqrt{\tilde{\Omega}_{h0}}}\sqrt{\Omega_{h}}\right),\label{eq:diffeq}
\end{equation}
where $'$ denotes the derivative with respect to $x =\ln a$. This
equation describes the evolution of dimensionless energy density of
dark energy. It can be solved exactly,
\begin{eqnarray}
\ln\Omega_{h}-2\ln\left(1+\alpha+2\frac{\sqrt{\Omega_{h}}}{\sqrt{\tilde{\Omega}_{h0}}}\right)
&-&\frac{(1+\alpha)}{2}\sqrt{\tilde{\Omega}_{h0}}\left[\ln(1-\sqrt{\Omega_{h}})-\ln(1+\sqrt{\Omega_{h}})\right]
+\frac{(1+\alpha)^2}{4}\tilde{\Omega}_{h0}\left[\ln(1-\Omega_{h})-\ln\Omega_{h}\right] \nonumber\\
&=&-(1+\alpha)\left[\frac{(1+\alpha)^2}{4}\tilde{\Omega}_{h0}-1\right]\ln
a +C_0,
\end{eqnarray}
where $C_0$ is an integration constant which can be obtained by
setting $a_0=1$
\begin{eqnarray}
C_0&=&\ln\Omega_{h0}-2\ln\left(1+\alpha+2\frac{\sqrt{\Omega_{h0}}}{\sqrt{\tilde{\Omega}_{h0}}}\right)
-\frac{(1+\alpha)}{2}\sqrt{\tilde{\Omega}_{h0}}\left[\ln(1-\sqrt{\Omega_{h0}})-\ln(1+\sqrt{\Omega_{h0}})\right]\nonumber\\
&+&\frac{(1+\alpha)^2}{4}\tilde{\Omega}_{h0}\left[\ln(1-\Omega_{h0})-\ln\Omega_{h0}\right].
\end{eqnarray}
It is obvious that $\Omega_{h}$ is a function of $\alpha$, $c$,
$\Omega_{h0}$ and $\tilde{\Omega}_{h0}$ (or $\omega$). Then, the
Friedmann equation (\ref{eq:EHFE}) is rewritten as
\begin{equation}
H^2=H^2_0\frac{\Omega_{m0}a^{-(3+\alpha)}}{1-\Omega_{h}}=H^2_0\frac{\Omega_{m0}(1+z)^{(3+\alpha)}}{1-\Omega_{h}},
\end{equation}
where $a_0/a=1+z$ is used in the second equal sign. From the
conservation equation of the holographic dark energy
(\ref{eq:DECE}), on has the equation of state (EoS) of holographic
dark energy
\begin{equation}
w_{h}=-1-\frac{1}{3}\frac{d \ln \rho_{h}}{d \ln
a}=-\frac{1}{3}\left(1+\alpha+\frac{2}{\sqrt{\tilde{\Omega}_{h0}}}\sqrt{\Omega_{h}}\right)=-\frac{1}{3}\left(1+\alpha+\frac{2}{c}\sqrt{\Omega_{h}}\right),\label{eq:DEEOS}
\end{equation}
where $w_{h}=p_{h}/\rho_{h}$.  The formula
$\rho_{h}=\frac{\Omega_{h}}{1-\Omega_{h}}\rho_{m0}a^{-3}$ and the
relation Eq. (\ref{eq:diffeq}) is used in the second equal sign. The
third equal sign is obtained by inserting the relation
$\tilde{\Omega}_{h0}=c^2$. From the above equation, one finds the
EoS of holographic dark energy is in the range of
\begin{equation}
-\frac{1}{3}\left(1+\alpha+\frac{2}{c}\right)<w_{h}<-\frac{1}{3}\left(1+\alpha\right),
\end{equation}
when one considers the holographic dark energy density ratio
$0\leq\Omega_{h}\leq1$.
Also, by using the Eq. (\ref{eq:RE}) and the assumption
$\Phi/\Phi_0=(a/a_0)^{\alpha}$, one obtains the deceleration
parameter as follows
\begin{equation}
q=-\frac{\ddot{a}}{aH^2}=\frac{1}{2}+\alpha+\frac{\alpha}{8+2\alpha}+\frac{6w_h\Omega_{h}}{4+\alpha}.\label{eq:dec}
\end{equation}
It is clear that the 'Standard' holographic dark energy will be
recovered in the limit $\alpha \rightarrow 0$. In Brans-Dicke theory
case, the dynamic behavior of the holographic dark energy is
determined by the parameters $c$ and $\alpha$. The holographic dark
energy can be quintessence, phantom and quitom as that in the
Standard case. The cosmic observational constraints on holographic
dark energy have been discussed by many authors
\cite{ref:ConHD1,ref:ConHD2,ref:ConHD3,ref:ConHD4,ref:ConHD5,ref:ConHD6,ref:ConHD7,ref:ConHD8}.
The joint analysis of SN, CMB shift parameter and BAO datasets, see
\cite{ref:ConHD8}, gives the results of the parameters in $1\sigma$
range: $c=0.91^{+0.26}_{-0.18}$ and $\Omega_{\rm m0}=0.29\pm 0.03$.
In this paper, in stead of giving any cosmic observational
constraints to the holographic dark energy in Brans-Dicke theory, we
are going to give some characteristic values of parameters to
describe the possible properties and evolutions of this kind of dark
energy. Obviously, with a positive value of $\alpha$, the range of
value the EoS of the holographic dark energy is enlarged. In Fig
\ref{fig:EOS}, the evolutions of EoS $w_{h}(z)$ and dimensionless
density parameter $\Omega_{h}(z)$ of the holographic dark energy and
the deceleration parameter $q(z)$ with respect to the redshift $z$
are plotted, where different parameter values $\Omega_{h0}=0.73$,
$c=0.31(c=0.91)$ and $\alpha=0.00005$ are adopted.
It is clear that an accelerated expansion of the universe is
obtained as shown in the bottom panels of Fig. \ref{fig:EOS}.
Now, one will take thought for variations of the Newton's constant
$G$ over the cosmological scale. Current constraints \cite{ref:VG}
on the variation of Newton's constant imply
\begin{equation}
\left|\frac{\dot{G}_{eff}}{G_{eff}}\right|\leq 10^{-11}yr^{-1},
\end{equation}
in our case, which corresponds to
\begin{equation}
\left|\frac{\dot{\Phi}}{\Phi}\right|=\alpha H \leq 10^{-11}yr^{-1}.
\end{equation}
It implies
\begin{equation}
\alpha \leq \frac{1}{H}\times 10^{-11}yr^{-1}.
\end{equation}
Considering the current value of Hubble constant
$h=0.73^{+0.03}_{-0.04}$ \cite{ref:CP2006}, one obtains the bounds
on $\alpha$, when the central value is taken
\begin{equation}
\alpha \leq 0.136438.
\end{equation}
Obviously, $\alpha=0.00005$ is under these bounds.

\section{Solution of scalar field $\Phi$ in Holographic Dark Energy Dominated Epoch}

Now, we study the scalar field solution in holographic dark energy
dominated epoch, under the assumptive solution
$\Phi/\Phi_0=(a/a_0)^\alpha$. In the holographic dark energy
dominated epoch, the dimensionless energy density of holographic
dark energy is $\Omega_{h}\approx 1$, and the dark matter energy
density can be neglected, i.e. $\rho_{m}\approx 0$. Also, in this
epoch, the holographic dark energy density equation (\ref{eq:EHHDE})
can be rewritten as
\begin{equation}
\rho_{h}=\frac{3c^2\Phi}{R^2_{eh}}=3H^2\Phi.\label{eq:rdrhoh}
\end{equation}
So, the evolution equation (\ref{eq:phiE}) of the scalar field
$\Phi$ is reduced to
\begin{eqnarray}
\ddot{\Phi}+3H\dot{\Phi}&=&\frac{\rho_{h}-3p_{h}}{2\omega+3}\nonumber\\
&=&\beta H^2\Phi,\label{eq:RdphiE}
\end{eqnarray}
where $\beta=\frac{\alpha^2+2\alpha+2\alpha/c}{4+\alpha}$ is a
constant, and the second equal sign is obtained by considering the
EoS (\ref{eq:DEEOS}) of the holographic dark energy and Eq.
(\ref{eq:rdrhoh}). Substituting the assumptive solution of
$\Phi/\Phi_0=(a/a_0)^\alpha$ into the above equation, one has
\begin{equation}
\alpha\eta\ddot{\eta}+\left(\alpha^2+2\alpha-\beta\right)\dot{\eta}^2=0,\label{eq:eta}
\end{equation}
where $\eta=a/a_0$ is used. This differential equation has the
solutions of power law and exponent. The power law solution is
\begin{equation}
\eta=C_1\left[\left(\alpha^2+3\alpha-\beta\right)t-C_2\alpha\right]^{\frac{\alpha}{\left(\alpha^2+3\alpha-\beta\right)}},
\end{equation}
where $C_1$ and $C_2$ are integral constant which are determined by
the initial condition $\eta_{0}=1$ and
$\dot{\eta}_{0}=\frac{\dot{a}_{0}}{a_0}=H_0$,
\begin{eqnarray}
1&=&C_1\left[\left(\alpha^2+3\alpha-\beta\right)t_0-C_2\alpha\right]^{\frac{\alpha}{\left(\alpha^2+3\alpha-\beta\right)}},\\
H_{0}&=&C_1\alpha\left[\left(\alpha^2+3\alpha-\beta\right)t_0-C_2\alpha\right]^{\frac{\alpha}{\left(\alpha^2+3\alpha-\beta\right)}-1},
\end{eqnarray}
here $t_{0}$ denotes the present time or the time when cold dark
matter can be neglected. To obtain a power law accelerated
expansion, one needs $\alpha/\left(\alpha^2+3\alpha-\beta\right)>1$,
i.e. $0<\alpha<\frac{-5+\sqrt{1+8/c}}{2}$ and $c<1/3$. When
$\alpha^2+3\alpha-\beta=0$ is respected, one has an exponential
solution
\begin{equation}
\eta=C_1 \exp(\lambda t),
\end{equation}
where $\lambda=H_0$ and $C_1=\exp(-\lambda t_0)$. Here, the case of
$\lambda<0$ is omitted, for its no accelerated properties. This
solution describes a de Sitter time-space, the result is consistent
with conventional consciousness. But in this case ($c\leq 1/5$),
$\alpha$ and $c$ has the algebraic relation
\begin{equation}
\alpha=\frac{-6+\sqrt{8/c-4}}{2},
\end{equation}
when $c\rightarrow 1/5$, one has $\alpha\rightarrow 0$, for example
$c=0.199998$, $\alpha=0.000017$. This is not surprising, because we
have assume the special solution $\Phi/\Phi_0=(a/a_0)^\alpha$. So,
in the limit of holographic dark energy dominated epoch, a de Sitter
like time-space can be obtained.

\section{Conclusions}

In this paper, the holographic dark energy model is explored in
Brans-Dicke theory where the holographic dark energy density
$\rho_{\Lambda} =3c^2 M^{2}_{pl} L^{-2}$ is replaced with
$\rho_{h}=3c^2 \Phi(t)L^{-2}$. Here $\Phi(t)=\frac{1}{8\pi G}$ is a
time variable Newton constant. With this replacement in Brans-Dicke
theory, it is found that no accelerated expansion universe will be
achieved when the Hubble horizon is taken as the role of IR cut-off.
When the event horizon takes the role of IR cut-off, an accelerated
expansion universe is obtained. In this case, the equation of state
of holographic dark energy $w_h$ takes in a modified form
$w_h=-\frac{1}{3}\left(1+\alpha+\frac{2}{c}\sqrt{\Omega_{h}}\right)$.
In the limit $\alpha\rightarrow 0$, the 'standard' holographic dark
energy is recovered. In the Brans-Dicke theory case of holographic
dark energy, the properties of the holographic dark energy is
determined by the parameter $c$ and $\alpha$ together. These
parameters would be obtained by confronting with cosmic
observational data. In stead of doing that, some characteristic
values of the parameters are given to describe the possible
properties and evolutions of the holographic dark energy in
Brans-Dicke case, see Fig. \ref{fig:EOS}. With this special solution
$\Phi/\Phi_0=(a/a_0)^\alpha$, one find power law and de Sitter like
time-space solutions in holographic dark energy dominated epoch.

\acknowledgements{This work is supported by NSF (10573003), NSF
(10703001), SRFDP (20070141034) and NBRP (2003CB716300) of P.R.
China. Also, L. X. Xu thanks X. Zhang for a valuable discussion.}

\end{document}